\documentclass[12pt]{iopart}
\expandafter\let\csname equation*\endcsname\relax
\expandafter\let\csname endequation*\endcsname\relax

\usepackage{amsfonts}
\usepackage{amsmath}
\usepackage{amssymb}
\usepackage{bm}
\usepackage{dcolumn}
\usepackage{graphicx}
\usepackage{graphics}
\usepackage[T1]{fontenc}
\usepackage[utf8]{inputenc}
\usepackage{latexsym}
\usepackage{rotating}
\usepackage{cite}
\usepackage[perpage,marginal]{footmisc}
\usepackage[colorlinks=true]{hyperref}
\usepackage[all]{hypcap}
\usepackage{xspace} % Sensible space treatment at end of simple macros
\usepackage[usenames]{color}
\usepackage{mathrsfs}
\usepackage{microtype}
\graphicspath{{fig/}}

\makeatletter
\renewcommand\footnoterule{%
  \kern-3\p@
  \hrule\@width2.5cm
  \kern2.6\p@}
\makeatother

\definecolor {darkgreen}{rgb}{0.2,0.7,0.2}

\newcommand\be{\begin{equation}}
\newcommand\ba{\begin{eqnarray}}
\newcommand\ee{\end{equation}}
\newcommand\ea{\end{eqnarray}}
\newcommand\bw{\begin{widetext}}
\newcommand\ew{\end{widetext}}

\newcommand{\nn}{\nonumber}

\newcommand{\GR}{{\mbox{\tiny GR}}}

\newcommand{\MAT}{{\mbox{\tiny mat}}}

\newcommand{\CS}{{\mbox{\tiny CS}}}

\newcommand{\NS}{*}

\newcommand{\mrm}{\mathrm}

\begin{document}
%\title{Toward a Better Understanding of How Compact Stars Approach Black Holes} 
%\title{Toward a Better Understanding of Universality as Compact Stars Approach Black Holes} 
\title{I-Love-Q Relations for Neutron Stars in dynamical Chern Simons Gravity} 

\author{Toral Gupta}
\address{Department of Physics, IIT Gandhinagar, Ahmedabad, India}

\author{Barun Majumder}
\address{Department of Physics, IIT Gandhinagar, Ahmedabad, India}

\author{Kent Yagi}
\address{Department of Physics, University of Virginia, Charlottesville, Virginia 22904, USA}
\address{Department of Physics, Princeton University, Princeton, New Jersey 08544, USA}

\author{Nicol\'as Yunes}
\address{eXtreme Gravity Institute, Department of Physics, Montana State University, Bozeman, Montana 59717, USA}

%\author{Anton B. Vorontsov}
%\affiliation{Department of Physics, Montana State University, Bozeman, MT 59717, USA.}

\date{\today}

%%%%%%%%%%%%%%%%%%%%%%%%%%%%%%%%%%%%%%%%%%%%%%%%%
\begin{abstract} 

%Testing GR with NSs
Neutron stars are ideal to probe, not only nuclear physics, but also strong-field gravity.
%Universal relations
Approximate universal relations insensitive to the star's internal structure exist among certain observables and are useful in testing General Relativity, as they project out the uncertainties in the equation of state.
%Example: I-Love-Q
One such set of universal relations between the moment of inertia $(I)$, the tidal Love number and the quadrupole moment $(Q)$ has been studied both in General Relativity and in modified theories. 
%focus of this paper: dCS
In this paper, we study the relations in dynamical Chern-Simons gravity, a well-motivated, parity-violating effective field theory, extending previous work in various ways.
%(I) Bounds from I-Love measurement
First, we study how projected constraints on the theory using the I-Love relation depend on the measurement accuracy of $I$ with radio observations and that of the Love number with gravitational-wave observations.
%Result 
Provided these quantities can be measured with future observations, we find that the latter could place bounds on dynamical Chern-Simons gravity that are six orders of magnitude stronger than current bounds.
%(II) I-Q and Q-Love in dCS
Second, we study  the I-Q and Q-Love relations in this theory by constructing slowly-rotating neutron star solutions to quadratic order in spin. 
%Result 
We find that the approximate universality continues to hold in dynamical Chern-Simons gravity, and in fact, it becomes stronger than in General Relativity, although its existence depends on the normalization of the dimensional coupling constant of the theory. 
%(III) Eccentricity profile
Finally, we study the variation of the eccentricity of isodensity contours inside a star and its relation to the degree of universality.  
%Result
We find that, in most cases, the eccentricity variation is smaller in dynamical Chern-Simons gravity than in General Relativity, providing further support to the idea that the approximate self-similarity of isodensity contours is responsible for universality.

\end{abstract}

%\pacs{04.30.Db,04.50Kd,04.25.Nx,97.60.Jd}

%04.30.Db Wave generation and sources
% 04.50.Kd Modified theories of gravity
% 04.25.-g Approximation methods; equations of motion
%04.25.Nx Post-Newtonian approximation; perturbation theory; related approximations
%97.60.Jd Neutron stars

\maketitle

%\tableofcontents

%%%%%%%%%%%%%%%%%%%%%%%%%%%%%%%%%%%
\section{Introduction}
\label{sec:intro}

%Importance of Testing GR
General Relativity (GR) has successfully passed all the tests performed so far in various regimes of gravity. Solar System experiments have placed very tight constraints on weak-field modifications to GR~\cite{TEGP,Will:LRR}. Various cosmological observations have placed bounds on large scale modifications~\cite{Jain:2010ka,Clifton:2011jh,Joyce:2014kja,Koyama:2015vza,Salvatelli:2016mgy}. The recent discovery of gravitational waves (GWs) from binary black hole (BH) coalescences by the LIGO Scientific Collaboration (LSC) and the Virgo Collaboration~\cite{Abbott:2016blz,Abbott:2016nmj,TheLIGOScientific:2016pea,Abbott:2017vtc,Abbott:2017oio} allowed one to probe gravity in both the strong and dynamical field regimes for the first time~\cite{TheLIGOScientific:2016src,Yunes:2016jcc}. Future GW observations of binary BH coalescences will significantly improve the ability of probing gravity in such a regime~\cite{Gair:2012nm,Yunes:2013dva,Berti:2015itd,Yagi:2016jml,Barausse:2016eii,Chamberlain:2017fjl}.

%Testing GR with NSs
Neutron star (NS) observations are also excellent tools to perform strong-field tests of gravity~\cite{psaltis-review,Stairs:2003yja,Wex:2014nva}. One very timely example is the detection of GWs and electromagnetic waves from a binary neutron star merger GW170817~\cite{TheLIGOScientific:2017qsa}. This observation allows for bounds on the propagation speed of GWs, on gravitational Lorentz violation and on the equivalence principle~\cite{Monitor:2017mdv,Hansen:2014ewa}. Unlike BH observations, one can use NS observations to probe how non-GR effects may arise in spacetimes with matter. For example, binary pulsar observations~\cite{Damour:1996ke,Damour:1998jk,freire,Shao:2017gwu} have been used to probe spontaneous scalarization in scalar-tensor theories~\cite{Damour:1992we,Damour:1993hw}, which arises due to the nonlinear coupling between matter and a scalar field, and is absent in BH spacetimes. Since the density inside a NS exceeds the nuclear saturation density, one can use independent measurements of NS quantities, such as the mass and radius, to probe the equation of state (i.e.~the relation between pressure and density) of supranuclear matter, which is currently unknown~\cite{lattimer_prakash2001,lattimer-prakash-review,Lattimer:2012nd,steiner-lattimer-brown,Ozel:2012wu,Miller:2013tca,Ozel:2015fia,Ozel:2016oaf}. Unfortunately, however, this means that there typically exists a degeneracy between equation-of-state effects and strong-field modifications in NS observations, rendering the latter unfeasible without knowledge of the former. 

%Universal Relations
One way to overcome such a problem is to use approximate universal relations among NS observables, i.e.~relations that do not depend strongly on the underlying equation of state (EoS), to project out uncertainties in nuclear physics and allow for tests of GR. Universal relations are known to exist for example between the oscillation frequencies of NSs and the stellar average density~\cite{andersson-kokkotas-1996,andersson-kokkotas-1998}, or between the binding energy of a star and its compactness~\cite{lattimer_prakash2001}. Recently, two of us found another example of universal relations, one that holds among dimensionless versions of the moment of inertia ($I$), the tidal Love number (Love) and the quadrupole moment ($Q$)~\cite{I-Love-Q-Science,I-Love-Q-PRD}. The universality, in fact, depends sensitively on how one adimensionalizes these NS observables~\cite{Majumder:2015kfa}. For example, the I-Q universality is lost for rapidly-rotating NSs if one fixes the angular velocity~\cite{doneva-rapid}, but it is restored if one fixes the dimensionless spin parameter instead~\cite{pappas-apostolatos,Chakrabarti:2013tca}. 

%%Why I-Love-Q
What is the origin of the I-Love-Q universality? Reference~\cite{Stein:2013ofa} argued that the fact that isodensity contours inside a star are approximately elliptically self similar is the reason. In other words, the fact that the eccentricity of such isodensity contours is approximately constant throughout the star is responsible for the approximate universality. Indeed, Ref.~\cite{Stein:2013ofa} showed that the NS eccentricity varies by only $\sim 20\%$ throughout its interior for all EoSs they considered, providing evidence for this explanation. Moreover, Ref.~\cite{Yagi:2014qua} showed that the relations become significantly less universal if one artificially forces the variation of the eccentricity of isodensity contours to increase. This explanation is also consistent with the results of~\cite{Sham:2014kea,Chan:2015iou}, who suggested the origin of the universality is associated with all realistic EoSs being ``similar'' to an incompressible (constant density star) one, in which case the elliptical isodensity approximation becomes exact.

%Testing GR with universal relations
The I-Love-Q relations have applications to strong-field tests of gravity, astrophysics~\cite{Newton:2016weo}, gravitational-wave physics~\cite{I-Love-Q-Science,I-Love-Q-PRD} and nuclear physics~\cite{Baubock:2013gna,Psaltis:2013fha} (see~\cite{Yagi:2016bkt,Doneva:2017jop} and references therein for detailed and recent reviews), but let us concentrate on the former. Future radio observations may provide the first measurements of the moment of inertia of the primary pulsar in J0737-3039~\cite{lattimer-schutz,kramer-wex}, while future gravitational-wave observations may provide the first measurements of the NS tidal Love number~\cite{read-markakis-shibata,flanagan-hinderer-love,hinderer-lackey-lang-read,damour-nagar-villain,lackey,lackey-kyutoku-spin-BHNS,Read:2013zra,Hotokezaka:2016bzh}\footnote{The recent GW170817 detection already placed upper bounds on the tidal Love number with the NS mass of 1.4$M_\odot$~\cite{TheLIGOScientific:2017qsa}, although these are quite weak.}. These two independent observations define a region in the I-Love plane, whose size is determined by the uncertainty in the observations. If GR holds, then the GR I-Love relation will pass through this region, thus placing constraints on any deviation. Modified theories that predict parametric deviations in the GR I-Love curve can then be constrained, as is the case in scalar tensor theories~\cite{Doneva:2014faa,Pani:2014jra,Doneva:2016xmf}, $f(R)$ gravity~\cite{Doneva:2015hsa}, Einstein-dilaton Gauss-Bonnet gravity~\cite{Kleihaus:2014lba,Kleihaus:2016dui}, dynamical Chern-Simons (dCS) gravity~\cite{I-Love-Q-Science,I-Love-Q-PRD} and Eddington-inspired Born-Infeld gravity~\cite{Sham:2013cya}. Whether such I-Love tests are more constraining than other existing tests, using for example Solar System observations, depends of course on how stringent the other constraints are.  

%dCS and NSs
In this paper, we study the I-Love-Q universal relations in dCS gravity~\cite{jackiw:2003:cmo,Smith:2007jm,CSreview}, a parity violating effective field theory of gravity. The effective action in this theory consists of the Einstein-Hilbert term and a Pontryagin density (i.e.~the contraction of the Riemann tensor and its dual) coupled to a dynamical scalar field with a standard kinetic term. Such a modification to the action arises naturally in string theory~\cite{polchinskiVol1,polchinskiVol2}, in loop quantum gravity upon the promotion of the Barbero-Immirzi parameter to a field~\cite{taveras,calcagni}, and in effective field theories of inflation~\cite{weinberg-CS}. Stellar solutions in dCS gravity have of course been studied before: non-rotating solutions do not acquire any modifications, while slowly-rotating solutions to first order in rotation were found in~\cite{yunes-CSNS} and to second order in rotation in~\cite{Yagi:2013mbt}, following the Hartle-Thorne formalism~\cite{hartle1967,hartlethorne}. 

We here extend previous results in several ways. First, we evaluate the quadrupole moment in dCS gravity following~\cite{Yagi:2013mbt}, and study the universality in the I-Q and Q-Love relations. Second, we investigate how the universality depends on the normalization of the coupling constant in dCS gravity. Third, we look at the eccentricity profile of isodensity contours inside a NS in dCS gravity to determine whether the elliptical isodensity explanation continues to hold in this theory. Lastly, we investigate the dependence of the projected bounds on dCS gravity using the I-Love relation on the measurement accuracy of the moment of inertia and the tidal Love number.

%%%%%%  
\begin{figure}[htb]
\centering
\includegraphics[width=9.5cm,clip=true]{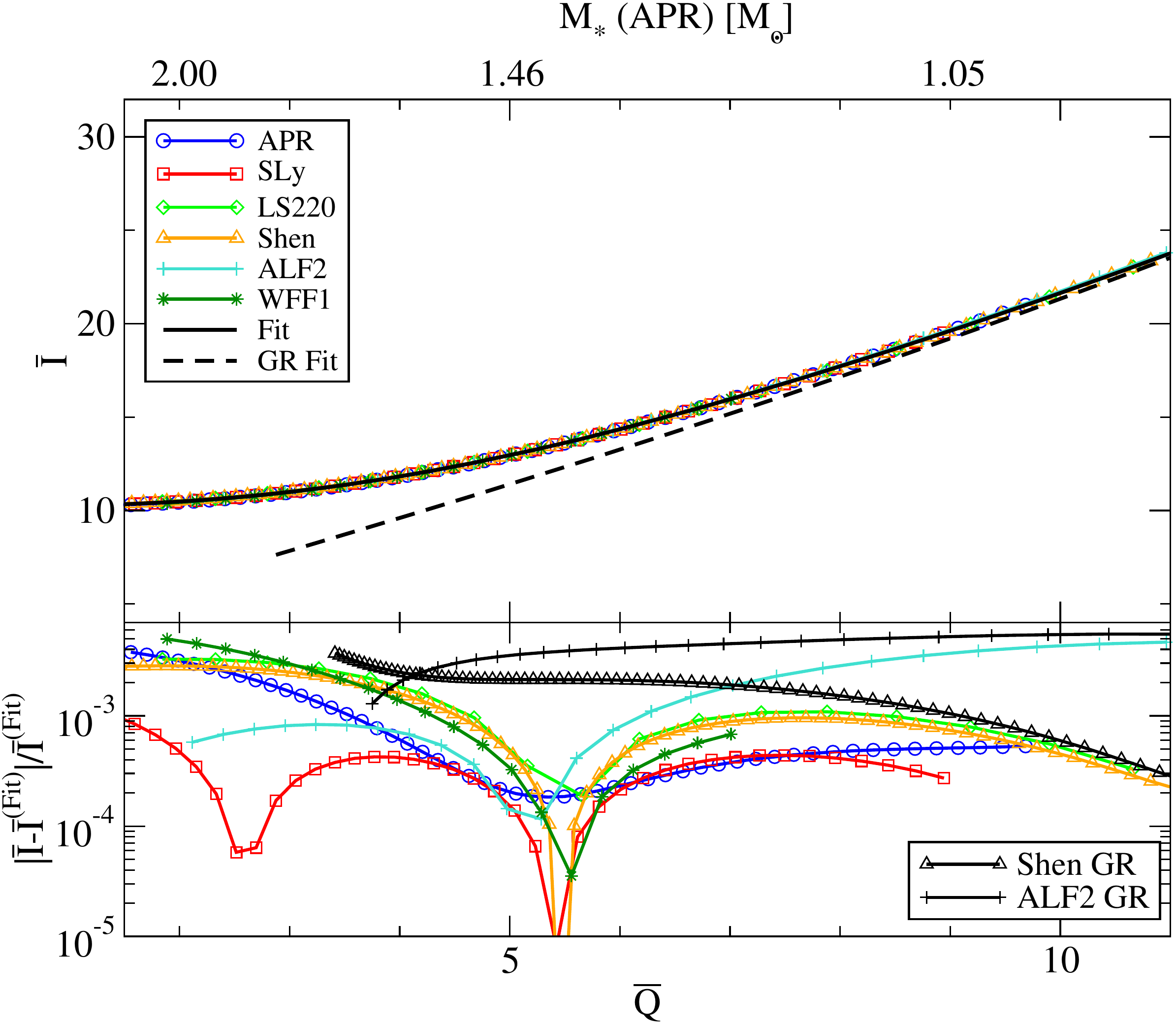}
\caption{\label{fig:I-Q} (Color Online) (Top) Universal relation between the dimensionless moment of inertia $\bar I$ and quadrupole moment $\bar Q$ for various realistic EoSs in dCS gravity, using a dimensionless coupling constant normalized by the NS mass and set to $\bar \xi = 10^3$. We also show a global fit to all dCS I-Q curves (black solid), together with a global fit to all I-Q curves in GR~\cite{I-Love-Q-Science,I-Love-Q-PRD} (black dashed). Observe how the relation deviates from the GR one as one decreases $\bar Q$ (as one considers more massive and compact NSs). (Bottom) Fractional difference between any one I-Q curve in dCS gravity and the global fit, together with the fractional difference between two I-Q relations in GR and a global fit in GR using the Shen and ALF2 EoSs. Observe that the amount of universality in dCS gravity is similar to that in GR for large or small $\bar Q$, while the dCS universality is stronger when $\bar Q \sim 5$. }
\end{figure}
%%%%%%  

%----------------------------------------------
\subsection*{Executive Summary}

%I-Q
The first topic we investigate is the universality of the I-Love-Q relations. The top panel of Fig.~\ref{fig:I-Q} presents the I-Q relation in dCS gravity for various EoSs, where we adimensionalize the dCS coupling constant by the NS mass. Observe that the relation is still approximately EoS universal, with dCS deviations from the GR I-Q relation becoming more prominent for more compact (smaller $\bar Q$) stars. The bottom panel shows the fractional difference between any one I-Q curve in dCS and a global fit constructed from all dCS data sets, together with the EoS variation found in GR for two EoSs for comparison. Observe that the EoS variation in dCS gravity is similar to that in the GR case, for large and small $\bar Q$. Observe also that when  $\bar Q \sim 5$, the EoS variation in dCS gravity is roughly one order of magnitude smaller than that in GR; this is because when $\bar Q$ is large (small), the EoS variation is dominated by the GR (dCS) contribution, while the dCS and GR contributions partially cancel each other when $\bar Q \sim 5$ for the specific value of the dimensionless coupling constant we chose in this example. 

%Normalization
The second topic we investigate is how the degree of universality changes when one chooses different normalizations for the dCS coupling constant. As we showed in Fig.~\ref{fig:I-Q}, the degree of universality in dCS gravity is similar to (and in some cases even stronger than) the degree of universality in the GR case, when one adimensionalizes the coupling constant by the NS mass. We find, however, that if one adimensionalizes the coupling constant with either the curvature length of the system, or simply with a solar mass, the universality is lost. These results demonstrate that, even in dCS gravity, the existence of universality in the I-Love-Q relations depends sensitively on how one adimensionalizes the dimensional quantities of the problem. We expect this to remain true in other theories of gravity that depend on dimensional coupling constants, such as in Einstein-dilaton Gauss-Bonnet gravity. 

%Eccentricity profile
The third topic we investigate is whether the self-similar elliptical isodensity explanation of the I-Love-Q universality persists in dCS gravity. We find that as one increases the dCS coupling, the eccentricity variation throughout the star becomes smaller \emph{and} the universality becomes stronger, thus supporting the self-similar elliptical isodensity explanation~\cite{Stein:2013ofa,Yagi:2014qua}. However, if one increases the dCS coupling too much, the EoS variation is dominated by the dCS contribution, and then, although the eccentricity variation continues to decrease, the universality does not necessarily improve. This suggests that the origin of the universality in modified theories of gravity may be more complicated than in GR, as additional non-GR contributions come into play.

\begin{figure}[hbtp]
\begin{center}
\includegraphics[width=9.5cm,clip=true]{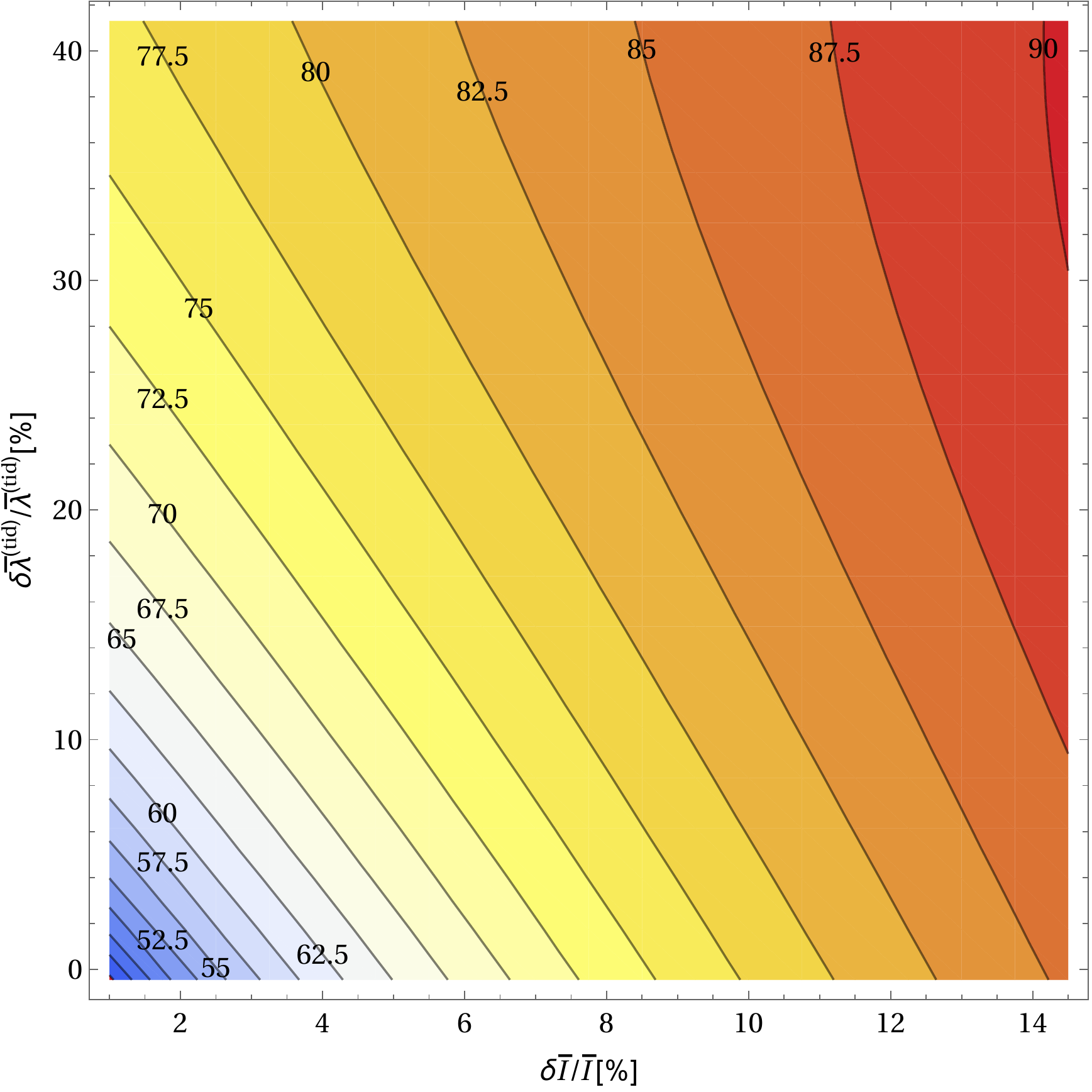}
\caption{\label{fig:alpha_constraint}~(Color Online) Contour plot of projected upper bounds on the dCS coupling constant $\sqrt{\alpha}$ in km as a function of the fractional measurement accuracy of the NS moment of inertia (horizontal axis) and tidal Love number (vertical axis). We assume that the former is measured from future radio observations of binary pulsars, while the latter is measured from future GW observations of NS binaries. Observe that the constraints are always six orders of magnitude stronger than current bounds ($\sqrt{\alpha} < \mathcal{O}(10^8)$km) from Solar System~\cite{alihaimoud-chen} and table-top~\cite{kent-CSBH} experiments, approximately independently of the accuracy to which these quantities can be measured.}
\end{center}
\end{figure}

With the universality of the I-Love-Q relations established and explored, we then investigate the dependence of I-Love constraints on dCS on the accuracy to which the moment of inertia and the Love number could be measured by binary pulsar and gravitational wave observations. Figure~\ref{fig:alpha_constraint} shows how the bounds on the dCS coupling constant varies with the measurement accuracy. Observe that the bounds are $\mathcal{O}(10^2)$km, namely six orders of magnitude stronger than current Solar System~\cite{alihaimoud-chen} and table top~\cite{kent-CSBH} experiments. This is because NS observations allow one to probe stronger gravitational field regimes, where the dCS corrections would be naturally larger than those present in the weak field regime. Observe also that these constraints only depend on these measurements being possible, with the strength of the constraint not strongly dependent of the accuracy to which these quantities are measured. 
%%%%%%  
%%%%%%  

The remainder of this paper presents the details of the results summarized above and it is organized as follows. Section~\ref{sec:review} briefly reviews dCS gravity and explains a few approximations that we adopt throughout the paper. Section~\ref{sec:matt-space-decomp} describes the perturbation scheme that we use to construct slowly-rotating NS solutions in the theory by extending the Hartle-Thorne formalism. Section~\ref{sec:I-L-Q} reviews the structure of the field equations and how we numerically obtain dCS corrections to physical observables, such as the moment of inertia, the quadrupole moment, the tidal Love number and the stellar eccentricity. We also explain how we adimensionalize the moment of inertia, tidal Love number and quadrupole moment and derive dCS corrections for such dimensionless quantities. We further comment on how to evaluate the stellar eccentricity. Section~\ref{sec:num-res} presents all of our numerical results. Section~\ref{sec:disc} provides a final summary of our work and describes possible avenues for future work. 

%%%%%%%%%%%%%%%%%%%%%%%%%%
\section{Review of Dynamical Chern-Simons gravity}
\label{sec:review}

In this section, we review dCS gravity and explain the approximations that we adopt throughout the paper. We mostly follow~\cite{Yagi:2013awa,Yagi:2013mbt}.

%------------------------------------------------------------------
\subsection{Action and Field Equations}
\label{sec:action}
We begin by presenting the action in dCS gravity.
The modification to the Einstein Hilbert action is introduced by adding the Pontryagin density coupled to a dynamical scalar field $\vartheta$ to the Lagrangian, namely~\cite{CSreview}
\ba
S &\equiv & \int d^4x \sqrt{-g} \Big[ \kappa_g \mathcal{R} + \frac{\alpha}{4} \vartheta \, \mathcal{R}_{\nu\mu \rho \sigma} {}^* \mathcal{R}^{\mu\nu\rho\sigma}  - \frac{\beta}{2} \nabla_\mu \vartheta \nabla^{\mu} \vartheta + \mathcal{L}_{\MAT} \Big]\,.
\label{action}
\ea 
Here, $\mathcal{R}$ is the Ricci scalar, $\mathcal{L}_{\MAT}$ is the matter Lagrangian density, $g$ is the determinant of the metric, $\kappa_g \equiv 1/(16\pi)$, and $\alpha$ and $\beta$ are coupling constants of the theory. The dual of the Riemann tensor $^*\mathcal{R}^{\mu\nu\rho\sigma}$ is here defined by
\be
{}^* \mathcal{R}^{\mu\nu\rho\sigma} \equiv \frac{1}{2} \varepsilon^{\rho \sigma \alpha \beta} \mathcal{R}^{\mu\nu}{}_{\alpha \beta}\,,
\ee
where $\varepsilon^{\rho \sigma \alpha \beta}$ is the Levi-Civita tensor. For simplicity, we have neglected any potential for the scalar field. We take $\vartheta$ and $\beta$ to be dimensionless, which forces $\alpha$ to have dimensions of (length)$^{2}$~\cite{yunespretorius,kent-CSBH}. Solar System~\cite{alihaimoud-chen} and table-top~\cite{kent-CSBH} experiments place bounds on the characteristic length scale of the theory, $\xi_\CS^{1/4} < \mathcal{O}(10^8)$ km~\cite{alihaimoud-chen,kent-CSBH} where
\be
\label{xi-def}
\xi_\CS \equiv \frac{\alpha^2}{\beta \kappa_g}\,.
\ee

We next look at the field equations derived from the action in Eq.~\eqref{action}. The modified Einstein equations are given by 
\be
G_{\mu\nu} + \frac{\alpha}{\kappa_g} C_{\mu\nu} =\frac{1}{2\kappa_g} (T_{\mu\nu}^\mrm{mat} + T_{\mu\nu}^\vartheta)\,.
\label{field-eq}
\ee
Here, $G_{\mu\nu}$ is the Einstein tensor and $T_{\mu\nu}^\mrm{mat}$ is the matter stress-energy tensor. The C-tensor and the scalar field stress-energy tensor in Eq.~\eqref{field-eq} are given by
\begin{align}
C^{\mu\nu} & \equiv  (\nabla_\sigma \vartheta) \epsilon^{\sigma\delta\alpha(\mu} \nabla_\alpha \mathcal{R}^{\nu)}{}_\delta + (\nabla_\sigma \nabla_\delta \vartheta) {}^* \mathcal{R}^{\delta (\mu\nu) \sigma}\,, \\
\label{eq:Tab-theta}
T_{\mu\nu}^\vartheta & \equiv  \beta (\nabla_\mu \vartheta) (\nabla_\nu \vartheta) -\frac{\beta}{2} g_{\mu\nu} \nabla_\delta \vartheta \nabla^\delta \vartheta\,.
\end{align}
The dynamical scalar field $\vartheta$ satisfies the evolution equation
\be
\square \vartheta = -\frac{\alpha}{4 \beta} \mathcal{R}_{\nu\mu \rho \sigma} {}^*\mathcal{R}^{\mu\nu\rho\sigma}\,.
\label{scalar-wave-eq}
\ee
Taking the divergence of the field equation~\eqref{field-eq}, and using the Bianchi identity and Eq.~\eqref{scalar-wave-eq}, we find that the matter stress-energy tensor obeys the same conservation law as in GR:
\be
\nabla_{\nu} T_\MAT^{\mu\nu}=0\,.
\label{mat-cons}
\ee

%------------------------------------------------------------------
\subsection{Slow-rotation and Small-coupling Approximation}
Throughout this paper, we work within the slow-rotation approximation. We assume that the dimensionless spin parameter of a NS satisfies 
\be
\chi \equiv  \frac{J}{M_*^{2}}  \ll 1\,, 
\ee 
where $M_*$ is the NS mass and $J \equiv |\vec{J}|$ is the magnitude of the spin angular momentum.
This is an excellent approximation for old NSs, such as binary pulsars and NSs in a binary that are about to coalesce, one of the main sources of GWs for ground-based detectors. We estimate NS quantities in dCS gravity within this slow-rotation approximation, keeping terms up to ${\cal{O}}(\chi^{2})$.
 
Throughout this paper we also work in the small-coupling approximation. For isolated NSs, this approximation requires that the dimensionless coupling 
\be
\label{zeta-def}
\zeta \equiv \frac{\xi_\CS  M_\NS^2}{R_*^6} \ll 1\,, 
\ee
where $R_*$ and $M_\NS$ are the NS radius and mass, $\sqrt{R_*^3/M_\NS}$ is the curvature length scale of the star, and $\xi_\CS$ is defined in Eq.~\eqref{xi-def}. This requirement ensures that the dCS modification term (second term) in the action in Eq.~\eqref{action} is always much smaller than the Einstein-Hilbert term. This, in turn, allows us to treat dCS gravity as an effective theory, so that we can safely neglect possible higher-order terms in the action, and allowing the theory to easily reduce to GR in the low energy limit. When calculating NS quantities in dCS gravity we will work in the small-coupling approximation to ${\cal{O}}(\zeta)$.

One may wonder whether the small-coupling approximation is consistent with the current constraint on dCS gravity. Normalizing the small-coupling condition to neutron stars, one can rewrite Eq.~\eqref{zeta-def} as 
\begin{align}
\xi_\CS^{1/4} &\ll 25 \; {\textrm{km}} \; \left(\frac{M_*}{1.4 M_\odot}\right) \left(\frac{0.18}{C}\right)^{3/2}\,,
\end{align} 
where $C \equiv M_*/R_*$ is the stellar compactness. Observe that such a requirement is much more stringent than current bounds from Solar System~\cite{alihaimoud-chen} and table-top~\cite{kent-CSBH} experiments, $\xi_\CS^{1/4} \leq \mathcal{O}(10^8)$ km, mentioned earlier~\cite{alihaimoud-chen,kent-CSBH}. Therefore, requiring the small-coupling approximation is clearly not in conflict with current bounds on dCS gravity. 

Given these approximations, all solutions we obtain will be \emph{bivariate expansions}, i.e.~expansions in two independent small parameters, $\chi$ and $\zeta$. Therefore, it will be natural to first decompose any quantity $A$ as follows
\be
A = A^\GR + \alpha'{}^2\;A^\CS\,,
\ee
and then to further decompose the GR and dCS pieces as a sum over $\chi$ to ${\cal{O}}(\chi^{2})$, where $\alpha'$ is a book-keeping parameter that labels the half-order in the small-coupling approximation, i.e.~${\cal{O}}(\zeta) = {\cal{O}}(\alpha'^{2})$. Composing both expansions, we can typically write any quantity $A$ as 
\begin{align}
\label{eq:decomp}
A = \sum_{k=0}^{k=2} \sum_{\ell=0}^{\ell=2} \chi'^{k} \alpha'^{\ell} A_{(k,\ell)}\,,
\end{align}
where $\chi'$ is another book-keeping parameter that labels the order in the slow-rotation approximation.

%%%%%%%%%%%%%%%%%%%%%%%%%%
\section{Ansatz for Slowly-Rotating Neutron Stars}
\label{sec:matt-space-decomp}

In this section we first describe the metric ansatz and the stress-energy tensor for matter that we use to construct slowly-rotating NS solutions in dCS gravity. 

%------------------------------------------------------------------
\subsection{Metric Ansatz and Bivariate Expansion} 

We begin by considering the following ansatz for the metric, which corresponds to a generic stationary and axisymmetric metric in the Hartle-Thorne coordinates~\cite{hartle1967}:
\begin{align}
ds^2 &= -e^{\bar{\nu}(r)} \left[1+2 \bar{h}(r,\theta) \right] dt^2 + e^{\bar{\lambda}(r)} \left[ 1+\frac{2\bar{m}(r,\theta)}{r-2\bar{M}(r)} \right] dr^2 \nn \\
&  + r^2 \left[ 1+2\bar{k}(r,\theta) \right] \left\{ d\theta^2 + \sin^2 \theta \left[ d\phi - \bar{\omega}(r,\theta) dt \right]^2 \right\}\,.
\label{metric-ansatz-rth}
\end{align}
Here $\bar{\nu}$ and $\bar{\lambda}$ are functions on $r$ only that represent spin-independent contributions, while $\bar{h}$, $\bar{k}$, $\bar{m}$ and $\bar{\omega}$ are functions of $(r,\theta)$ and correspond to spin corrections. The quantity 
\be
\bar{M}(r) \equiv \frac{\left( 1-e^{-\bar{\lambda}(r)} \right)r}{2}
\ee
measures the enclosed mass within the radius $r$. 

Since we are treating rotation perturbatively, we need to be careful with the coordinate system we work in~\cite{hartle1967}. When considering rotating stars in polar coordinates ($r$,$\theta$), the fractional change in quantities like the energy density $\bar \rho$ and pressure $\bar p$ does not remain small near the surface. Thus, we adopt new coordinates ($R,\Theta$), first proposed by Hartle~\cite{hartle1967}, such that the energy density in the new radial coordinate $\rho(R)$ in the rotating configuration is the same as that in the non-rotating configuration:
\be
\bar \rho \left[ r(R,\Theta ), \Theta \right] = \rho (R) = \rho^\mrm{(non-rot)}(R), \quad \Theta = \theta\,.
\ee
Since the energy density and pressure are connected via an EoS, it follows that the pressure also does not acquire any spin corrections in the new coordinate system:
\be
\rho(R) = \rho^\mrm{(non-rot)}(R) \; \rightarrow \; p(R)=p^\mrm{(non-rot)}(R)\,.
\ee 
Moreover, since we treat rotation perturbatively and the changes in the density and pressure enter at ${\cal{O}}(\chi^{2})$, we can introduce the new coordinates perturbatively via
\be
r(R,\Theta) = R + \xi(R,\Theta)\,,
\ee
where $\xi = {\cal{O}}(\chi^{2})$. 

In the new coordinates, the line element becomes
\ba
ds^2 &=& - \left[\left( 1+2h+\xi \frac{d \nu}{dR} \right)e^\nu - R^2 \omega^2 \sin^2 \Theta \right]dt^2 \nn \\
& & -2 R^2 \omega \sin^2 \Theta dt d\phi + \left[ R^2 (1+2k) + 2 R \xi \right] \sin^2\Theta d\phi^2 \nn \\
& & + e^\lambda \left(1 +\frac{2m}{R-2M} + \xi \frac{d\lambda}{dR} + 2 \frac{\partial \xi}{\partial R} \right) dR^2 \nn \\
& & + 2e^\lambda \frac{\partial \xi}{\partial \Theta} dRd\Theta + \left[ R^2 (1+2k) + 2R\xi \right]d\Theta^2\,, 
\label{metric-ansatz-RTh}
\ea
where we have retained terms only up to ${\cal{O}}(\chi^{2})$ and we have defined
\be
M(R) \equiv \frac{\left( 1-e^{-\lambda (R)} \right)R}{2}\,,
\label{M-lambda}
\ee
with the stellar mass $M_* = M(R_*)$. Bivariately expanding the above line element, we then find that all metric functions and the coordinate function $\xi$ can be expanded as in Eq.~\eqref{eq:decomp}: 
\begin{align}
\nu (R) &= \nu_{(0,0)} (R)\,,  \\
\lambda (R) &= \lambda_{(0,0)} (R)\,,  \\ 
\omega (R,\Theta) &= \chi' \omega_{(1,0)} (R,\Theta)+\alpha'^2 \chi' \omega_{(1,2)} (R,\Theta) + \mathcal{O}(\chi'^3) \,,  \\
A (R,\Theta) &= \chi'^2 A_{(2,0)}(R,\theta)+ \alpha'^2 \chi'^2 A_{(2,2)} (R,\Theta)+ \mathcal{O}(\chi'^4) \,, 
\end{align}
with $A = (h,k,m,\xi)$. 

Let us make several observations about the above decomposition. First, the metric functions $\nu$ and $\lambda$ do not contain dCS corrections because non-rotating NS solutions in dCS gravity are the same as those in GR due to parity protection~\cite{jackiw:2003:cmo,Yunes:2007ss,CSreview}. Second, also because of parity considerations, only the $(t,\phi)$ component of the metric contains terms odd in $\chi'$. Third, terms of $\mathcal{O}(\alpha')$ do not appear in the metric because of the structure of the field equations, which force the scalar field $\vartheta$ to be proportional to $\alpha$ in the small-coupling approximation~\cite{Yagi:2013mbt}. The above decomposition is therefore the minimal line element for a slowly-rotating star in dCS gravity within the small-coupling approximation. 

%------------------------------------------------------------------
\subsection{Matter Stress-energy Tensor and EoS}

We assume we can describe the matter inside the NS as a perfect fluid, so that its stress-energy tensor is given by
\be
T_{\mu\nu}^\MAT = (\rho + p ) u_\mu u_\nu + p \; g_{\mu\nu}\,,
\ee
where $u^\mu$ is the fluid's four-velocity given by
\be
u^\mu = (u^0, 0,0,\Omega_\NS u^0)\,.
\ee
The quantity $\Omega_\NS$ is the \emph{constant} angular velocity of the fluid, and thus of the entire NS, since we model stars in uniform rotation. The normalization condition of the four-velocity $u_\mu u^\mu = -1$ determines $u^0$ in terms of the metric functions~\cite{Yagi:2013mbt}.

To close the system of equations, one needs to specify the EoS that relates $p$ and $\rho$. In this paper, we consider six different EoSs: APR~\cite{Akmal:1998cf}, SLy~\cite{Douchin:2001sv}, Lattimer-Swesty with nuclear incompressibility of 220 MeV (LS220)~\cite{LATTIMER1991331,OConnor:2009iuz}, Shen~\cite{Shen:1998gq,Shen:1998by,OConnor:2009iuz}, WFF1~\cite{Wiringa:1988tp} and ALF2~\cite{Alford:2004pf}. For the LS220 and Shen EoSs, we impose neutrino-less, beta-equilibrium conditions. The APR, SLy and WFF1 EoSs are relatively soft, while LS220 is intermediate and Shen is a stiff EoS. The ALF2 EoS is a hybrid between APR nuclear matter and color-flavor-locked quark matter.

%%%%%%%%%%%%%%%%%%%%%%%%%%
\section{Constructing Neutron Star Solutions}
\label{sec:I-L-Q}

In this section, we first review the structure of the field equations at each order in the dimensionless spin parameter. We then explain how we make each of the I-Love-Q quantities dimensionless following~\cite{I-Love-Q-Science,I-Love-Q-PRD} and derive dCS corrections to these quantities. We also derive the dCS correction to the stellar eccentricity in terms of the metric perturbation.

%------------------------------------------------------------------
\subsection{Structure of the Field Equations}
\label{sec:structure}

The field equations are first expanded order by order in both $\chi'$ and $\alpha'$ and then solved numerically also order by order. At any given order, the equations are first solved numerically in the stellar interior, imposing regularity at the center, and then analytically in the exterior region, imposing asymptotic flatness at spatial infinity. One then matches the two solutions at the stellar surface, the radial location where the pressure vanishes,  by requiring continuity and smoothness of all metric functions. From these metric functions, we then construct NS observables, such as the mass, moment of inertia and quadrupole moment. Below, we present a brief summary of the structure of the field equations at each order in spin; the full expressions can be found in~\cite{Yagi:2013mbt}.
 
%zeroth order in spin 
At zeroth order in spin, there is no dCS correction. The field equations reduce to the Einstein equations for a non-rotating configuration. Using the (t,t) and (R,R) components of the equations, together with the conservation of matter stress-energy tensor from Eq.~\eqref{mat-cons}, one finds the Tolman-Oppenheimer-Volkoff (TOV) equation. The stellar mass $M_*$ appears as an integration constant in the exterior solution.

%first order in spin
At first order in spin, the only non-vanishing part of the modified Einstein equations is the $(t,\phi)$ component, which can be decomposed into a GR and dCS part of $\mathcal{O}(\alpha'{^2})$. The dCS part depends on the scalar field, whose evolution equation forces it to be of $\mathcal{O}(\alpha')$. One can Legendre decompose both of these equations such that the set of partial differential equations becomes a set of ordinary differential equations. Boundary conditions at the stellar center and at spatial infinity make the $\ell=1$ mode the only non-vanishing piece of the equations. The magnitude of the angular momentum $J$ can be read off from the asymptotic behavior of the $g_{t\phi}$ metric component, and the moment of inertia is then found by 
\be
I \equiv \frac{J}{\Omega_*}\,.
\ee

%second order in spin
At second order in spin, one can again apply a Legendre decomposition to find that both the $\ell = 0$ and $\ell = 2$ modes contribute in the modified Einstein equations. Both modes have a GR and dCS contribution, with the latter again entering at $\mathcal{O}(\alpha'{^2})$. The $\ell=0$ mode gives a spin correction to the mass, which is irrelevant in this paper. We thus focus on the $\ell =2$ mode, which allows us to calculate the quadrupole moment $Q$ from the integration constants in the exterior solution. In particular, the gravitational potential $U$, which can be read off from the $g_{tt}$ component, has the asymptotic form\footnote{This definition of $Q$ is different from the one used in~\cite{Yagi:2013mbt} by a factor of 2. We choose to use this definition here so that the GR contribution matches the Geroch-Hansen multipole moments~\cite{geroch,hansen} and the moments used by Hartle and Thorne~\cite{hartlethorne}.} 
\be
\label{eq:UA_CS}
U^{\CS} = -\frac{3}{2}  \frac{Q}{R^{3}} \hat{S}^{i}\hat{S}^{j} n^{<ij>} + \mathcal{O}\left( \frac{M_*^4}{R^4} \right)\,,
\ee
where $\hat{S}^{i}$ is the unit spin angular momentum vector of a NS and $n^i$ is the unit vector that points to a field point.

The above solutions automatically lead to the mass and radius of the star $M_{*}$ and $R_{*}$, the moment of inertial $I$ and the quadrupole moment $Q$, but other important quantities can also be computed, such as the electric-type tidal Love number $\lambda^\mrm{(tid)}$. This quantity characterizes the quadrupolar tidal deformability of a star and it is defined as~\cite{hinderer-love}
\be
Q^\mathrm{(tid)} = - \lambda^\mathrm{(tid)} \, \mathcal{E}^\mathrm{(tid)}\,,
\ee
where $Q^\mathrm{(tid)}$ is the \emph{tidally-induced} quadrupole moment (not to be confused with the spin-induced quadrupole moment $Q$), while $\mathcal{E}^\mathrm{(tid)}$ is the external tidal field strength. One can treat tidal perturbations in a similar manner to perturbations due to spin. The main difference is that one must set $\omega=0$ (i.e. no parity-odd perturbation) in Eq.~\eqref{metric-ansatz-RTh} and not impose asymptotic flatness when solving the equations. Then, $Q^\mathrm{(tid)}$ and $\mathcal{E}^\mathrm{(tid)}$ can be read off from the coefficients of the $P_2(\Theta)/R^3$ and $P_2(\Theta) R^2$ parts of the metric perturbation $h$ when asymptotically expanded away from the NS surface, with $P_2$ the $\ell=2$ Legendre polynomial. Since dCS gravity modifies the odd-parity sector, $\lambda^\mathrm{(tid)}$ does not acquire any dCS correction~\cite{Yagi:2011xp}.

%------------------------------------------------------------------
\subsection{Dimensionless I-Love-Q and Eccentricity Profiles}

We next explain how we adimensionalize physical quantities.
Let us first focus on the moment of inertia. We make this quantity dimensionless via 
\begin{align}
\label{eq:Ibar}
\bar{I} \equiv \frac{I}{M_*^3} &= \frac{I_\GR+ \alpha'^2 I_\CS}{M_\GR^3}\,,
\end{align}
which we expand as
\begin{align}
\bar{I} = \bar{I}_\GR + \alpha'^2 \bar{I}_\CS\,,
\end{align}
where clearly
\begin{equation}
\bar{I}_\GR \equiv \frac{I_\GR}{M_\GR^3}\,, \quad
\label{eq:IGR}
\bar{I}_\CS \equiv \frac{I_\CS}{M_\GR^3}\,.
\end{equation}
We recall that the NS mass does not acquire dCS corrections to leading order in spin.

Next, we look at the dCS correction to the dimensionless quadrupole moment. The latter is defined by
\begin{align}
\label{eq:Qbar}
\bar{Q} \equiv - \frac{Q}{M_*^3 \chi^2} &= -\frac{Q_\GR+\alpha'^2 Q_\CS}{M_\GR^3(\chi_\GR+\alpha'^2 \chi_\CS)^2}\,,
\end{align}
which we expand as
\begin{align}
\bar{Q} = \bar{Q}_\GR + \alpha'^2 \bar{Q}_\CS + {\cal O}(\alpha'^4),
\end{align}
where we have defined
\begin{equation}
\bar{Q}_\GR \equiv -\frac{Q_\GR}{M_\GR^3 \chi_\GR^2}\,, \quad 
\label{eq:QbarCS}
\bar{Q}_{\CS} \equiv -\frac{ Q_{\CS}}{M_{\GR}^3 \chi_{\GR}^2} + 2 \bar{Q}_{\GR}\frac{I_{\CS}}{I_{\GR}}.
\end{equation} 
Here we used the relation $\chi_\CS/\chi_\GR = I_\CS/I_\GR$. The second term in $\bar{Q}_{\CS}$ above typically dominates the first term in the stellar solutions we consider in this paper. 

Let us finally focus on the Love number. We make this quantity dimensionless via
\begin{align}
\label{eq:lovebar}
 \bar{\lambda}^\mathrm{(tid)} \equiv \frac{\lambda^\mathrm{(tid)}}{M_*^{5}} = \frac{\lambda^\mathrm{(tid)}_{\GR}}{M_{\GR}^{5}}\,,
\end{align}
where the last equality follows from the fact that the electric-type tidal deformability is not modified in dCS gravity as already explained. 

Before presenting numerical results, let us also discuss the eccentricity contours inside the star at a constant radius. Following~\cite{hartlethorne}, the eccentricity can be defined via
\begin{align}
e(R)=\sqrt{-3\left(k_2(R) + \frac{\xi_2(R)}{R}\right)}\,,
\end{align}
which we expand as
\be
e = e_\GR + \alpha'{}^2 e_\CS + {\cal O}(\alpha'^4)\,,
\ee 
where we have defined
\begin{align}
e_{\GR}(R)&=\sqrt{-3\left(k_2^{\GR}(R) + \frac{\xi_2^{\GR}(R)}{R}\right)}\,,
\\ 
e_{\CS}(R) &=- \frac{1}{2}\sqrt{\frac{-3}{k_2^{\GR}(R) + \frac{\xi_2^{\GR}(R)}{R}}}\left(k_2^{\CS}(R)+ \frac{\xi_2^{\CS}(R)}{R}\right)\,.
\end{align}
Here $k_2^{\GR}$ and $\xi_2^{\GR}$ ($k_2^{\CS}$ and $\xi_2^{\CS}$) are the coefficients of the $\ell = 2$ modes of $k_{(2,0)}$ and $\xi_{(2,0)}$ ($k_{(2,2)}$ and $\xi_{(2,2)}$ ) in a Legendre decomposition. Adimensinalizing this quantity, one further finds
\begin{align}
\frac{e(R)}{\chi} &= \frac{e_{\GR}+\alpha'^2 e_{\CS}}{\chi_{\GR}+\alpha'^2 \chi_{\CS}} \nonumber \\
&= \frac{e_{\GR}}{\chi_{\GR}} +\alpha'^2 \left( \frac{ e_{\CS}}{\chi_{\GR}} - \frac{e_{\GR}}{\chi_{\GR}} \frac{I_{\CS}}{I_{\GR}} \right) + {\cal O}(\alpha'^4)\,.
\end{align}

%%%%%%%%%%%%%%%%%%%%%%%%%%
\section{Numerical Results}
\label{sec:num-res}

In this section we present numerical results obtained by solving the field equations order by order, and calculating  the dimensionless quantities defined in the previous section. We conclude this section with a discussion of projected bounds we will be able to place on the theory given future observations of the moment of inertia and the tidal deformability.

%------------------------------------------------------------------
\subsection{I-Love-Q relations}

%%%%
\begin{figure*}[htb]
\begin{center}
\includegraphics[width=7.5cm,clip=true]{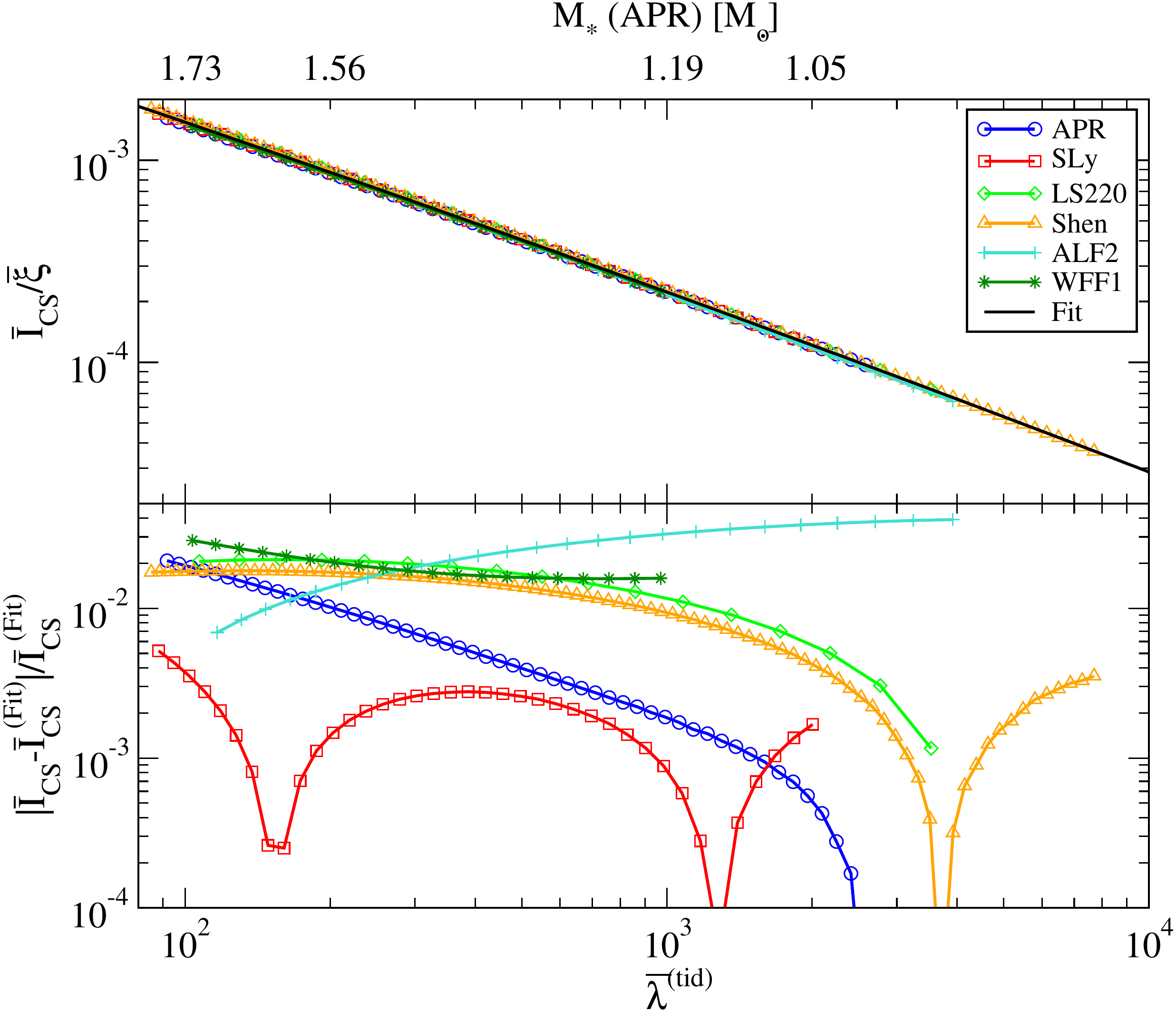}  \quad
\includegraphics[width=7.5cm,clip=true]{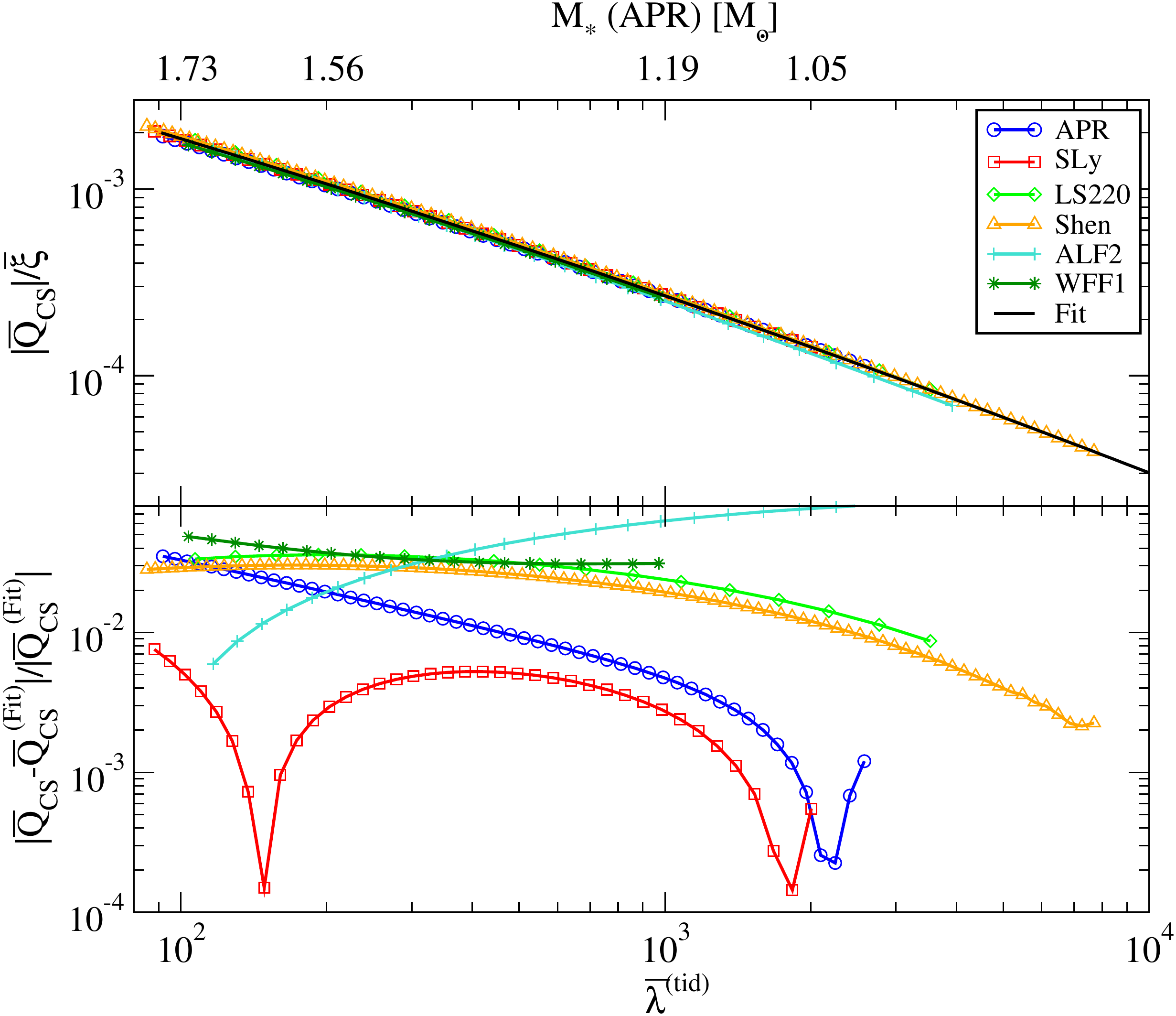}   
\caption{\label{fig:IQ-Love-CS-correction-only} (Color Online) (Top) dCS corrections to the I-Love and Q-Love relations, normalized to $\bar \xi$, for various EoSs and using the global fit of Eq.~\eqref{fitting-func}. The top axis shows the corresponding NS mass for a given $\bar \lambda^\mrm{(tid)}$ assuming the APR EoS in GR. (Bottom) Relative fractional difference between the data and the global fit, which represents the EoS variation in the relations. Observe that the universality holds to better than $5\%$.
}
\end{center}
\end{figure*}
%%%%

%I-Love, Q-Love CS curve explanation
We begin by studying dCS corrections to the I-Love and Q-Love relations, which we present in the top panels of Fig.~\ref{fig:IQ-Love-CS-correction-only}. We recall that $\bar \lambda^\mrm{(tid)}$ does not acquire any dCS corrections, and hence corrections to the I-Love-Q relations originate from changes to $\bar I$ and $\bar Q$. In Fig.~\ref{fig:IQ-Love-CS-correction-only}, we normalize $\bar I_\CS$ and $\bar Q_\CS$ with 
\be
\bar \xi \equiv \frac{\xi_\CS}{M_*^4}\,,
\ee
so that the dependence on the dCS coupling constant is factored out from corrections to the I-Love-Q relations in this figure. Observe that the dCS correction to the relations remains universal with respect to various EoSs. To estimate the degree to which these relations remain universal, we construct a global fit (shown by the black solid curves in the figure) for all the numerical data points in the form
\be
\label{fitting-func}
y_i = \exp \left[a_i + b_i(\log x_i) +c_i(\log x_i)^{2} \right]\,,
\ee
where the coefficients are summarized in Table~\ref{Fitting-params}. The bottom panel of Fig.~\ref{fig:IQ-Love-CS-correction-only} shows the fractional difference between each numerical data point and the fit, which corresponds to the EoS variation in these relations. Observe that the relations are universal to within $5\%$.
%%%%
\begin{table} 
\begin{center}
 \begin{tabular}{ m{1.5cm}  m{1.5cm}  m{1.5cm}  m{1.5cm}  m{1.5cm}}
 \hline \hline
   $y_i$ & $x_i$ & $a_i$ & $b_i$ & $c_i$ \\
   \hline
   $\bar{I}_\CS/\bar \xi$ & $\bar{\lambda}^\mrm{(tid)}$ & -2.943 & -0.718  & -0.011 \\
   $\bar{Q}_\CS/\bar \xi$ & $\bar{\lambda}^\mrm{(tid)}$ & -3.443 & -0.550  & -0.023\\
 \hline \hline
 \end{tabular}
 \caption{\label{Fitting-params} Estimated numerical coefficients for the fitting formula in Eq.~\eqref{fitting-func} for dCS corrections to the I-Love and Q-Love relations.}
\end{center}
\end{table}
%%%%

%%%%
\begin{figure*}[htb]
\begin{center}
\includegraphics[width=7.5cm,clip=true]{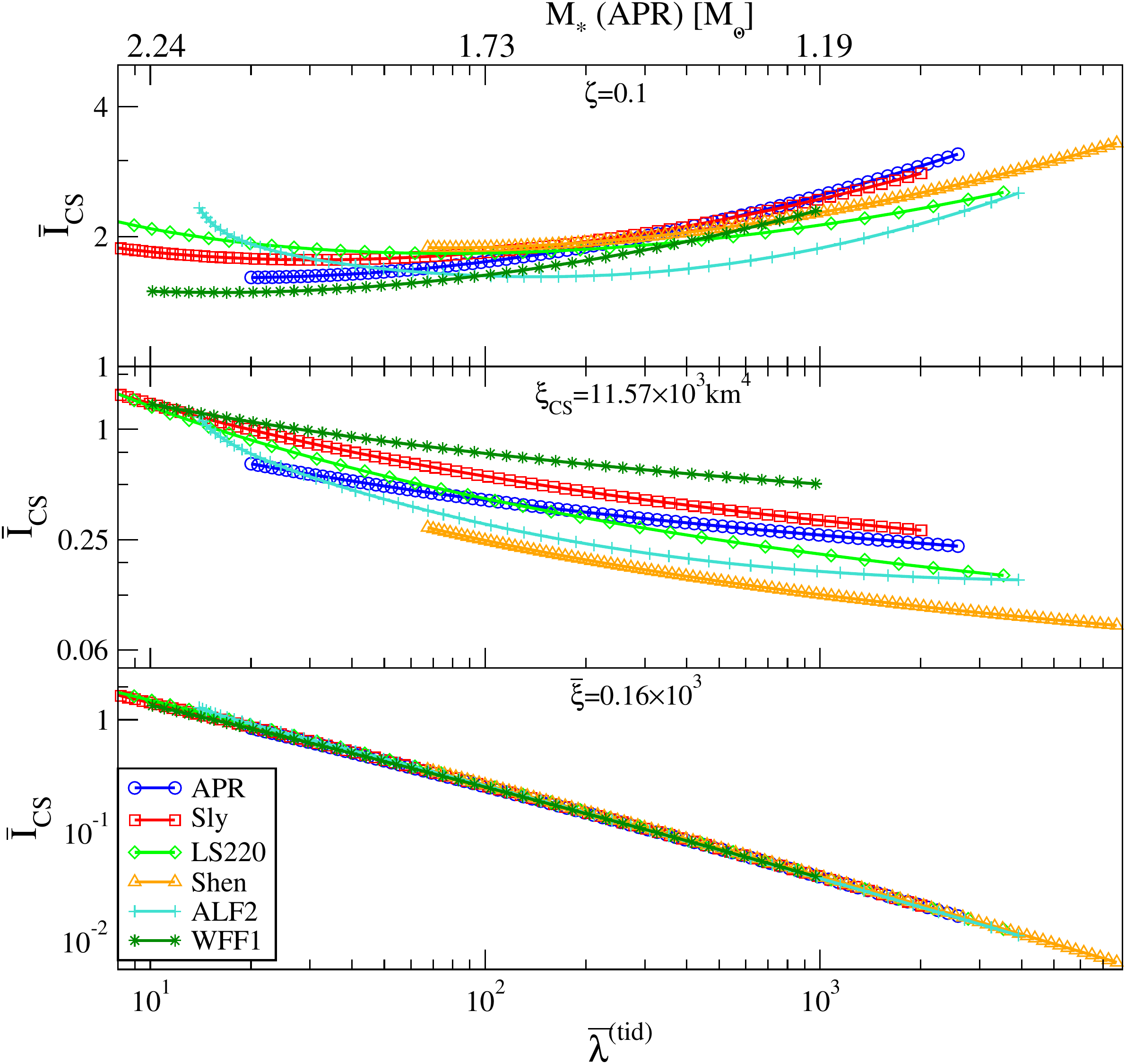}  \quad
\includegraphics[width=7.5cm,clip=true]{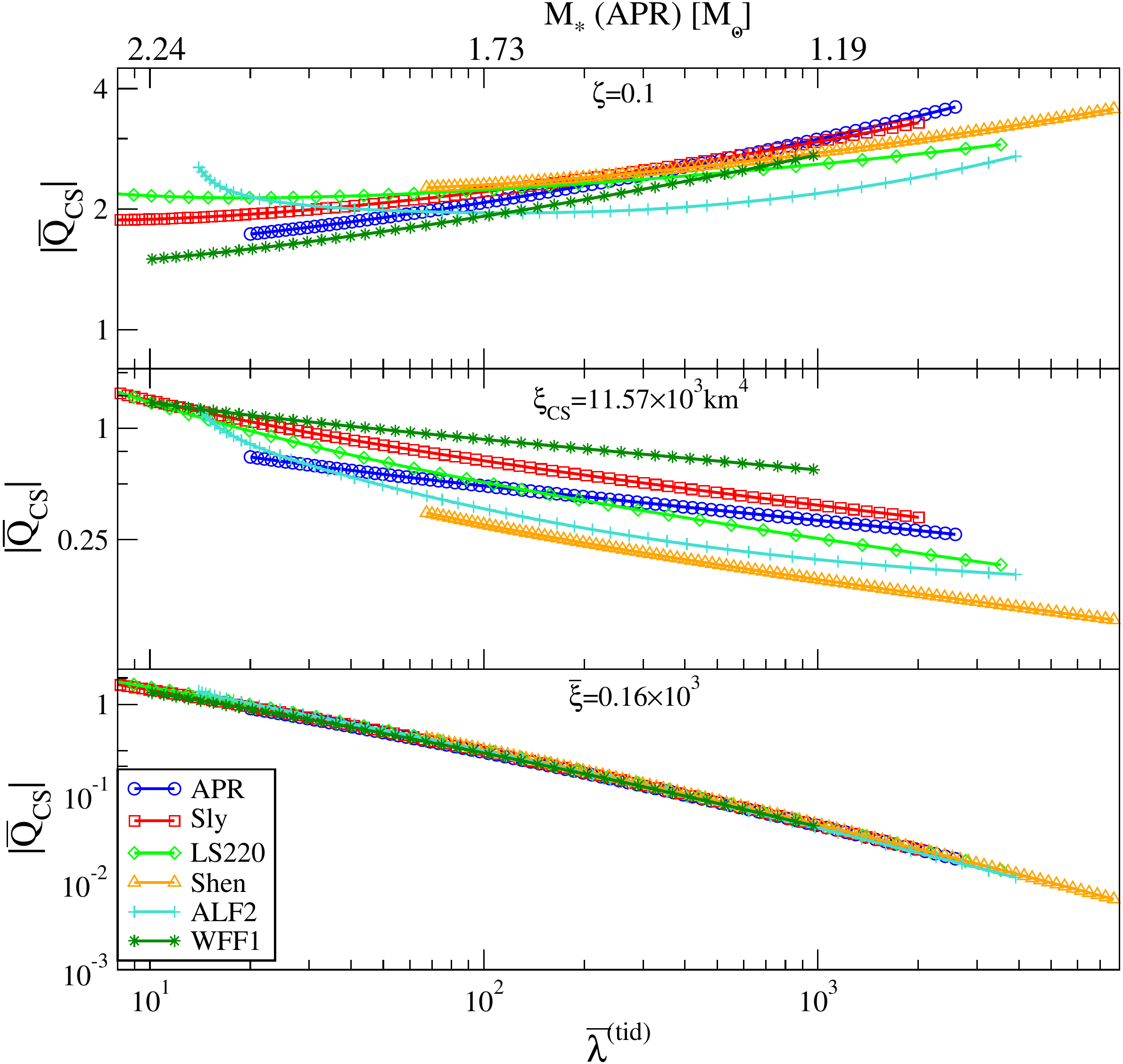}   
\caption{\label{fig:fixed_zeta_xiCS_xi} (Color Online) dCS corrections to the I-Love (left) and Q-Love (right) relations with $\zeta$ = 0.1 (top), $\xi_{\CS}$ = 11.57 $\times 10^{3}$ km$^4$ (middle) and $\bar{\xi}$= 0.16$\times 10^{3}$ (bottom). Observe that the relations remain universal only if one fixes $\bar \xi$.
}
\end{center}
\end{figure*}
%%%%

%Plot with different fixing parameter
Figure~\ref{fig:IQ-Love-CS-correction-only} corresponds to fixing the value of $\bar \xi$ (to unity), but how does the universality change if one normalizes the I-Love-Q relations differently? To address this question, Fig.~\ref{fig:fixed_zeta_xiCS_xi} presents the dCS corrections to the I-Love and Q-Love relations fixing $\zeta$ (top), $\xi_\CS$ (middle) and $\bar \xi$ (bottom), where the first two quantities are given in Eqs.~\eqref{zeta-def} and~\eqref{xi-def} respectively. We choose $\zeta=0.1$ to ensure the validity of the small-coupling approximation, and thus we choose values for $\xi_\CS$ and $\bar \xi$ that correspond to $\zeta=0.1$ for a NS with mass $2M_\odot$ and radius of $10$km. Observe that the universality is lost when the dCS correction is normalized by $\zeta$ or $\xi_\CS$, while it is recovered if one fixes $\bar \xi$. This clearly shows that whether the relations remain universal in modified theories of gravity depends crucially on how one normalizes observables with respect to the coupling constants of the theory.

%%%% 
 \begin{figure*}[htb]
\begin{center}
\includegraphics[width=7.5cm,clip=true]{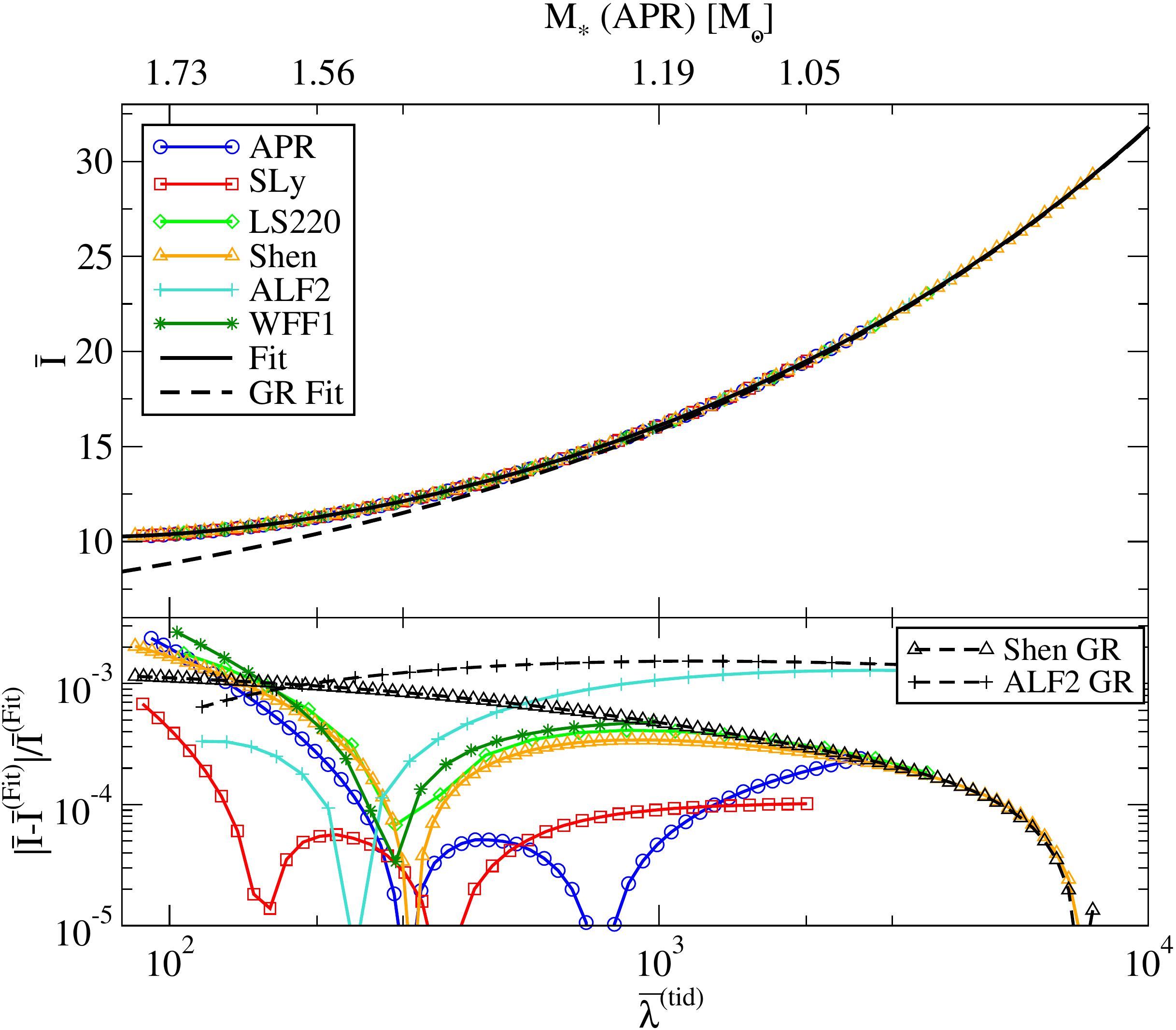}  \quad
\includegraphics[width=7.5cm,clip=true]{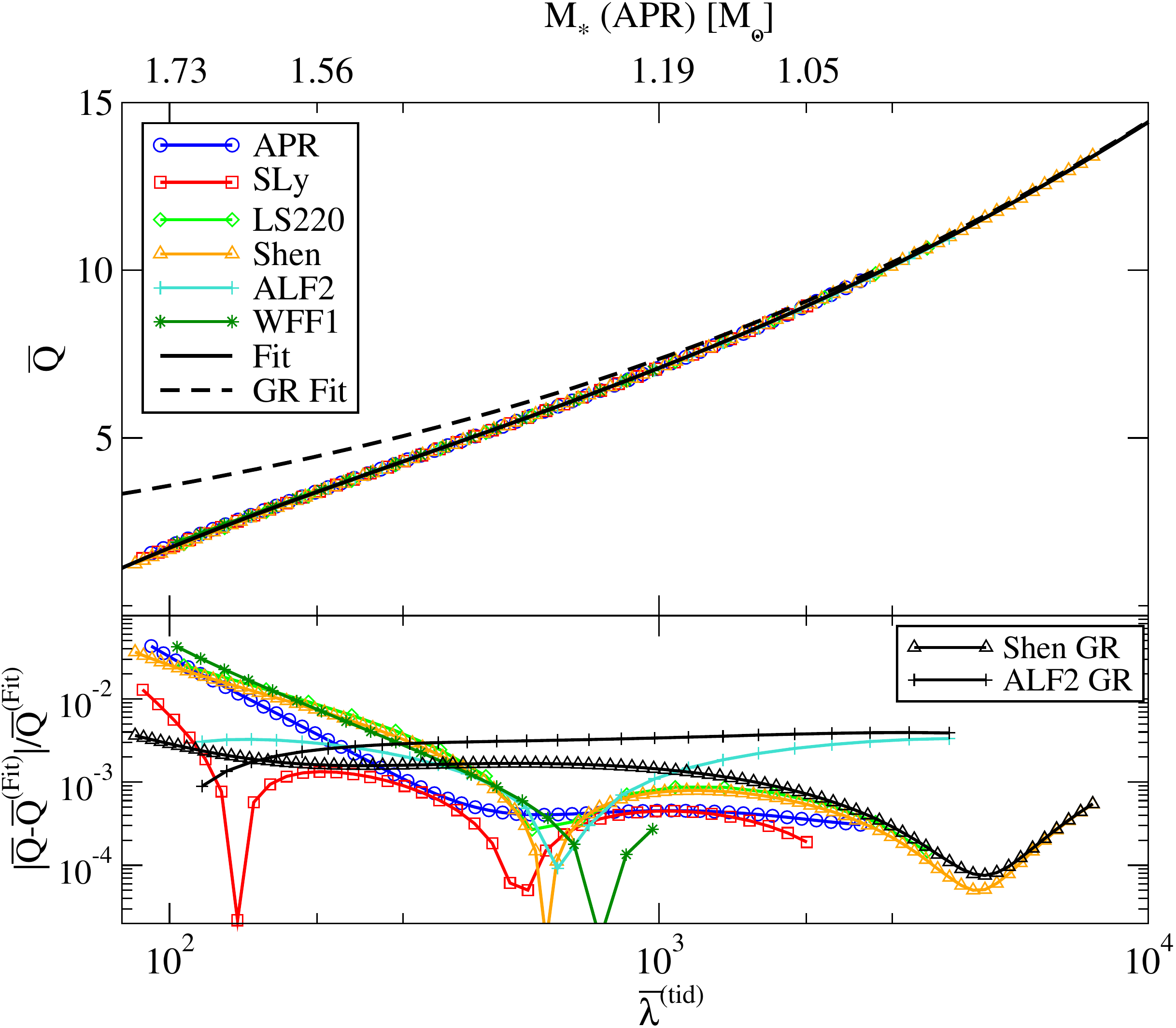}  
\caption{\label{fig:IL_QL} (Color Online) (Top) I-Love (left) and Q-Love (right) relations for various EoSs in dCS gravity with $\bar{\xi}=10^3$. The black dashed curve in each panel represents the best-fit relation in GR. (Bottom)  Fractional difference in each relation with respect to a global fit. For reference, we also present the fractional difference for the Shen and ALF2 EoSs in GR. Observe that the EoS variation in dCS gravity is comparable to that in GR for the I-Love relation, while the former is larger than the latter in the Q-Love relation for relativistic NSs with relatively small $\bar \lambda^\mrm{(tid)}$.}
\end{center}
\end{figure*}
%%%%

%I-Love, Q-Love relation
Having studied the dCS corrections to the I-Love-Q relations and which parameter one should fix to retain universality, we now study the full I-Love-Q relations in dCS gravity, including the GR contribution. Figures~\ref{fig:I-Q} and~\ref{fig:IL_QL} present such relations with $\bar \xi = 10^3$, together with the fractional difference from the global fit. Observe that the EoS variation is of $\mathcal{O}(0.1\%)$ for the I-Love and I-Q relations, and of $\mathcal{O}(1\%)$ for the Q-Love relation. We can compare this EoS variability with that present in GR, represented here through the fractional difference between the Shen and ALF2 EoSs relations in GR. Observe that in general, the EoS variation in dCS gravity is comparable to that in GR for the I-Love and I-Q relations, while the former is larger than the latter for the Q-Love relation, in particular for relativistic stars with smaller $\bar \lambda^\mrm{(tid)}$. Observe also that there is a certain parameter range for each relation in dCS gravity where the EoS variation is highly suppressed (e.g. $\bar \lambda^\mrm{(tid)} \sim 300$ for the I-Love relation). This is because the GR (dCS) contribution in the EoS variation dominates for large (small) $\lambda^\mrm{(tid)}$, while these two contributions partially cancel in the intermediate region.

%%%%
\begin{figure*}[htb]
\begin{center}
\includegraphics[width=7.5cm,clip=true]{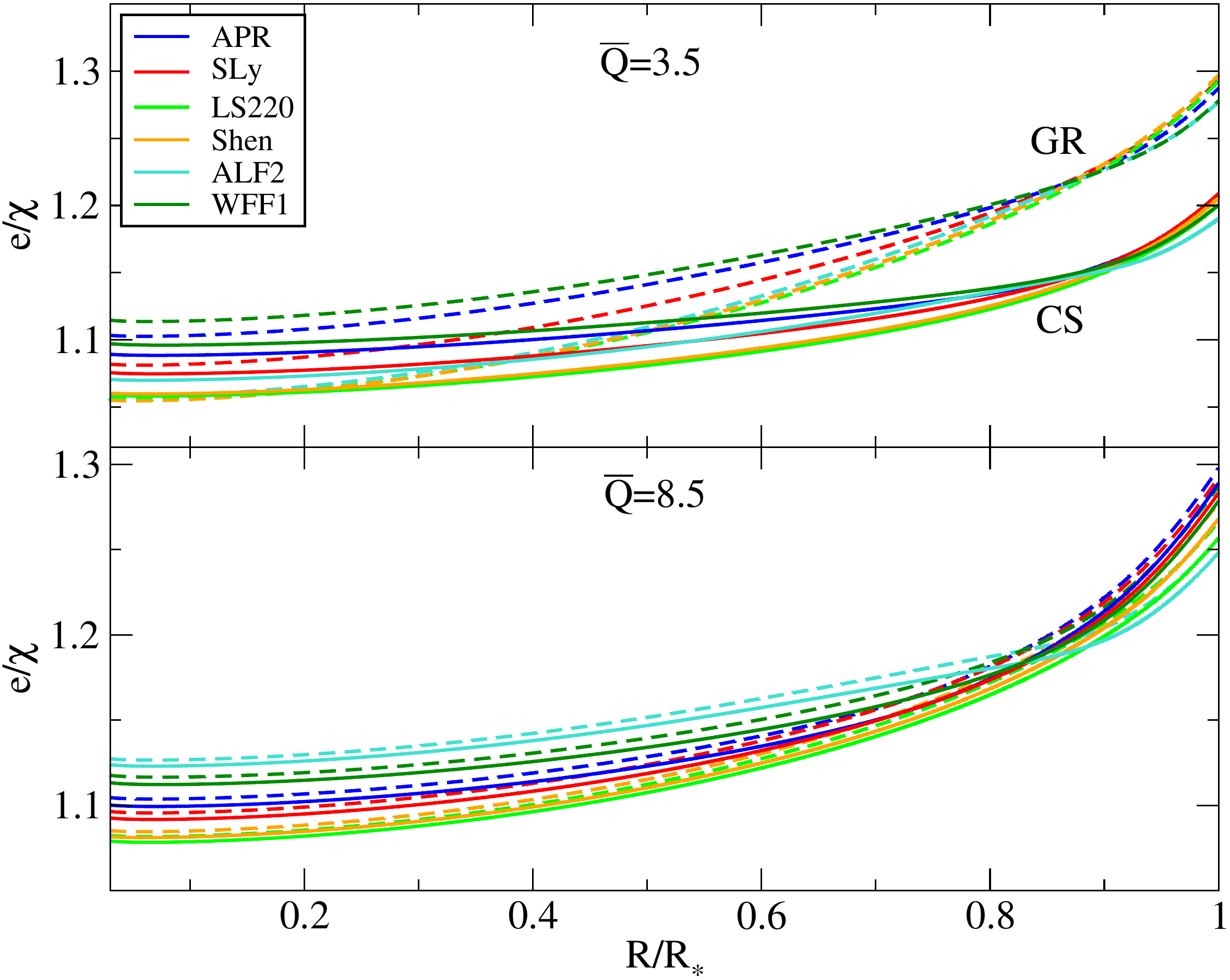}  \quad
\includegraphics[width=7.5cm,clip=true]{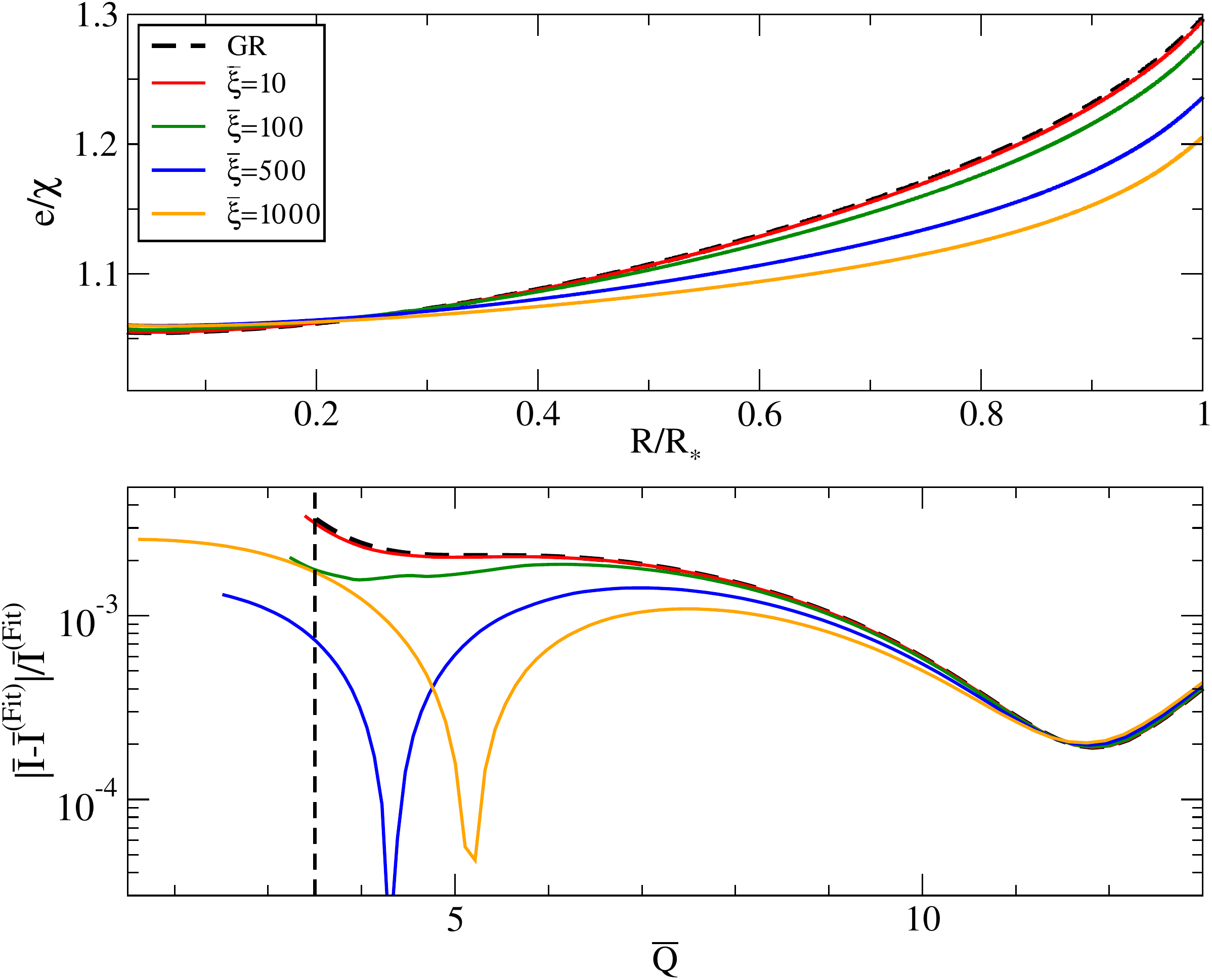}   
\caption{\label{fig:Eccentricity} (Color Online) (Left) Stellar eccentricity profile with $\bar{Q}=3.5$ (top) and $\bar{Q}=8.5$ (bottom) for various EoSs in GR and dCS gravity with $\bar \xi = 10^3$. (Top Right) Stellar eccentricity profile for the Shen EoS with $\bar Q = 3.5$ and various $\bar \xi$. Observe that the eccentricity variation throughout the star becomes smaller as one increases $\bar \xi$. (Bottom Right) Fractional difference from the fit in the I-Q relation for the Shen EoS and various $\bar \xi$. The vertical dashed line corresponds to $\bar Q = 3.5$. Observe that the EoS variation tends to become smaller for larger $\bar \xi$ where the eccentricity variation becomes also smaller. However, the EoS variation increases again if one increase $\bar \xi$ too much. 
 }
\end{center}
\end{figure*}
%%%%

%------------------------------------------------------------------
\subsection{Eccentricity Profile in dCS gravity}

Let us now study whether the elliptical isodensity explanation as the origin of the universality in GR~\cite{Yagi:2014qua} mentioned in Sec.~\ref{sec:intro} is still applicable in modified theories of gravity. First, we look at how the stellar eccentricity variation inside a star in dCS gravity changes from that in GR. The left panels of Fig.~\ref{fig:Eccentricity} present the eccentricity profile for various EoSs with $\bar Q = 3.5$ (top) and $\bar Q = 8.5$ (bottom) in GR and dCS with $\bar \xi = 10^3$. Observe that the eccentricity variation is smaller in dCS than in GR when $\bar Q$ is relatively small. On the other hand such a variation in dCS is almost indistinguishable from that in GR for relatively large $\bar Q$. This is because one approaches a Newtonian regime for larger $\bar Q$ values, in which dCS corrections are suppressed. The top right panel of Fig.~\ref{fig:Eccentricity} shows the eccentricity profile in dCS gravity in more detail. Here we choose the Shen EoS and $\bar Q = 3.5$, and study the eccentricity profile dependence on $\bar \xi$. One clearly sees that the variation becomes smaller for larger $\bar \xi$. \

We now investigate how such an eccentricity variation is related to the amount of EoS variation in the I-Love-Q relations. The bottom right panel of Fig.~\ref{fig:Eccentricity} presents the fractional difference in the I-Q relation for the Shen EoS from the fit for various values of $\bar \xi$. The vertical dashed line corresponds to $\bar Q = 3.5$. For this value of $\bar Q$, observe that the fractional difference becomes smaller as one increases $\bar \xi$ from 0 to 500, with which the eccentricity variation becomes also smaller as explained in the previous paragraph. This finding supports the conclusion in~\cite{Yagi:2014qua} that the I-Q relation (or relations among multipole moments) becomes stronger as the eccentricity variation becomes smaller. However, if one goes beyond $\bar \xi = 500$ (as done also e.g.~in Fig.~\ref{fig:IL_QL}), the EoS variation in the I-Q relation starts to increase, although the eccentricity variation keeps decreasing. This feature is opposite of what happens in GR and of what one would expect if the self-similarity of isodensity contours was solely responsible of the universality. This suggests that the origin of the universality in modified theories of gravity may be more complicated than that in GR due to the additional degrees of freedom present.

%------------------------------------------------------------------
\subsection{Future Observational Bounds}

%%%%%%  
\begin{figure}[hbtp]
\begin{center}
\includegraphics[width=9.5cm,clip=true]{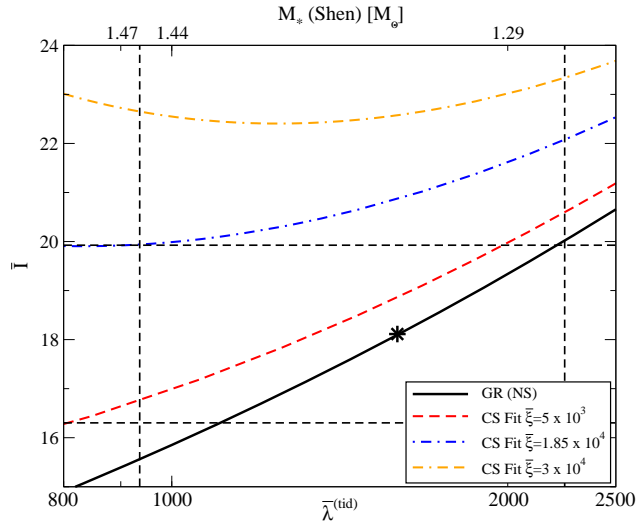}
\caption{\label{fig:GR_test} (Color Online) The best-fit I-Love relation in dCS gravity with various $\bar \xi$. We also show a 10\% measurement accuracy of the NS moment of inertia with future double binary pulsar observations, and a 40\% measurement accuracy of the NS tidal Love number with future gravitational wave observations, with a fiducial mass of $1.338M_\odot$ assuming the Shen EoS, which corresponds to the observed mass of the primary pulsar in J0737-3039. Observe that such observations, if realized, allow us to place the bound $\bar \xi < 1.85 \times 10^4$, which is six orders of magnitude more stringent than those from Solar System~\cite{alihaimoud-chen} and table-top~\cite{kent-CSBH} experiments in terms of the characteristic length scale $\xi_\CS^{1/4}$.
}
\end{center}
\end{figure} 
%%%%%%    

Universal relations project out uncertainties in nuclear physics, and thus they are very useful in probing strong-field gravity with NS observations. Let us then begin by reviewing how one can use these relations to probe dCS gravity following~\cite{I-Love-Q-Science,I-Love-Q-PRD}.  Let us imagine that we measure the moment of inertia of the primary pulsar (whose mass is estimated as $1.338 M_\odot$~\cite{burgay,lyne,kramer-double-pulsar}) in the double binary pulsar J0737-3039 to 10\% accuracy~\cite{lattimer-schutz,kramer-wex}. Let us also assume that we measure the tidal Love number of a $\sim 1.338 M_\odot$ NS\footnote{Even if we detect GWs from a NS binary with component masses different than the mass of the primary pulsar in J0737-3039, one can still measure the tidal Love number of a $1.338 M_\odot$ NS by Taylor expanding the tidal Love number as a function of the NS mass about $M_*=1.338 M_\odot$ and measuring the leading-order coefficient, as done in~\cite{messenger-read,delpozzo,Agathos:2015uaa,Yagi:2015pkc,Yagi:2016qmr}.} with an accuracy of 40\%~\cite{I-Love-Q-Science,I-Love-Q-PRD}. Figure~\ref{fig:GR_test} presents these errors in the I-Love plane as dashed lines with the fiducial values taken to be a $1.338 M_\odot$ NS (black star) assuming the Shen EoS (which gives the most conservative bound on dCS gravity among all the EoSs considered in this paper). The black solid curve corresponds to the I-Love relation in GR. 

The relation in modified theories of gravity has to pass through the error box in order to be consistent with the above hypothetical measurements. In Fig.~\ref{fig:GR_test}, we show the I-Love relation in dCS gravity with three different choices of $\bar \xi$. Observe that the relation with $\bar \xi < 1.85 \times 10^4$ is consistent with the measurement errors, thus ruling out $\bar \xi > 1.85 \times 10^4$. Using $M_*=1.338 M_\odot$, one can map such a bound to a constraint on the characteristic length scale $\sqrt{\alpha} \lesssim 86$ km, which is six orders of magnitude stronger~\cite{I-Love-Q-Science,I-Love-Q-PRD} than the bounds from Solar System~\cite{alihaimoud-chen} and table-top~\cite{kent-CSBH} experiments. Such a large improvement on the bound is realized because NS observations allow us to probe the strong-field regime, where dCS corrections become naturally large. One can also check that the above bound corresponds to $\zeta < 0.1$ in terms of the dimensionless coupling constant, and thus satisfies the small-coupling approximation. This finding shows the impact of using universal relations on probes of strong-field gravity.

We now go one step further and study how this putative bound on dCS changes with different measurement accuracy of the moment of inertia $\delta \bar I$ and tidal Love number $\delta \bar \lambda^\mrm{(tid)}$. The contours in Fig.~\ref{fig:alpha_constraint} present the bounds on $\sqrt{\alpha}$ as a function of the measurement accuracy $\delta \bar I / \bar I$ and $\delta \bar \lambda^\mrm{(tid)} / \bar \lambda^\mrm{(tid)}$. Observe that the bounds are all of $\mathcal{O}(10^2)$km and are not very sensitive to $\delta \bar I$ or $\delta \bar \lambda^\mrm{(tid)}$. This is because to map from bounds on $\bar{\xi}$ to bounds on $\sqrt{\alpha}$ one must take a fourth root, $\sqrt{\alpha} \sim \bar{\xi}^{1/4}$, which then softens the dependance of the $\sqrt{\alpha}$ constraint on $\delta \bar I$ and $\delta \bar \lambda^\mrm{(tid)}$. One should therefore be able to place bounds on dCS gravity that are roughly six orders of magnitude stronger than current ones if the moment of inertia and tidal Love number are measured with future NS observations, irrespective of the precise details of the measurement accuracy. 

%%%%%%%%%%%%%%%%%%%%%%%%%%
\section{Discussion}
\label{sec:disc}

In this paper we studied the I-Love-Q relations in dCS gravity. We found that whether such relations remain universal depends on what coupling parameter one fixes. If one fixes $\bar \xi$, the relations are universal, while the universality is lost for fixed $\zeta$ or $\xi_\CS$. By fixing $\bar \xi$, we found that the I-Love and I-Q relations are universal to $\mathcal{O}(0.1\%)$, which is comparable to that in GR. On the other hand, the Q-Love relation is universal to $\mathcal{O}(1\%)$, which is larger than in the GR case. We next studied whether the elliptical isodensity explanation as the origin of the universality still holds in dCS gravity. We found that the eccentricity variation inside a star in dCS is smaller that that in GR, and the universality in the I-Q relation becomes stronger as one increases $\bar \xi$. However, if one increases $\bar \xi$ too much, the universality becomes weaker while the eccentricity variation keeps decreasing. This suggests that the origin of the universality in non-GR theories may be more complicated than in GR.  Finally, we studied how one can use the I-Love relation to probe strong-field gravity. We found that future radio and gravitational wave observations should be able to place bounds on dCS gravity that are roughly six orders of magnitude stronger than current bounds, irrespective of the details of the measurement accuracy of the moment of inertia and tidal Love number.

One obvious direction for future work is to study the I-Love-Q relations for rapidly-rotating NSs. This could be done by modifying the publicly-available code RNS~\cite{stergioulas_friedman1995}. Such an extension is important to e.g. apply the I-Q relations to rapidly-spinning NS observations using NICER. Another avenue for future work includes studying other universal relations, such as those among higher-order multipole moments~\cite{Yagi:2014bxa} and various tidal Love numbers~\cite{Yagi:2013sva}. In particular, we studied parity-even tidal Love numbers in this paper but it would be interesting to consider parity-odd ones~\cite{damour-nagar,binnington-poisson,Pani:2015nua,Landry:2015zfa,Landry:2015cva,Delsate:2015wia,Landry:2015snx}, as these quantities would acquire non-vanishing dCS corrections. One can also study the universal relations and the origin of the universality in theories other than dCS gravity. For example, universal relations have not been studied within Lorentz-violating theories of gravity, such as Einstein-\AE ther~\cite{Jacobson:2000xp,Eling:2004dk,Jacobson:2008aj} and khronometric~\cite{Blas:2010hb} gravity.
In these theories, non-rotating~\cite{eling-AE-NS} and slowly-moving~\cite{Yagi:2013qpa,Yagi:2013ava} NS solutions have been constructed but slowly-rotating NS solutions have not been studied in the literature yet. Work along this direction is currently in progress.

%%%%%%%%%%%%%%%%%%%%%%%%%%%%%%%%%%%%%%%%%%
\section{Acknowledgements}
K.Y. acknowledges support from Simons Foundation and NSF grant PHY-1305682. N.Y. acknowledges support from NSF CAREER grant PHY-1250636 and NASA grants NNX16AB98G and 80NSSC17M0041. Some calculations used the computer algebra-systems \textsc{MAPLE}, in combination with the \textsc{GRTENSORII} package~\cite{grtensor}.

%%%%%%%%%%%%%%%%%%%%%%%%%%%%%%%%%%%%%%%%%%%%
%%%%%%%%%%%%%%%%%%%%%%%%%%%%%%%%%%%%%%%%%%%%
\section*{References}

\bibliographystyle{iopart-num}
\bibliography{citations}
\end{document}